\documentstyle[12pt]{article}
\def\lromn#1{\uppercase\expandafter{\romannumeral#1}}

\begin{document}


\begin{flushright}
TU/95/493 \\
TUM-HEP-229/95 \\
SFB-375/24 \\
hep-ph/9511381
\end{flushright}

\vspace{12pt}

\begin{center}
\begin{Large}

\renewcommand{\thefootnote}{\fnsymbol{footnote}}
\bf{
Particle Production and Gravitino Abundance\\
after Inflation}
\footnote[1]
{Work supported in part by the Grant-in-Aid for Science Research from
the Ministry of \\
\hspace*{0.6cm} Education, Science and Culture of
Japan No. 06640367, and by the Sonderforschungsbereich \\
\hspace*{0.6cm} 375-95:
Research in Particle-Astrophysics of the Deutsche Forschungsgemeinschaft
and \\
\hspace*{0.6cm}
the ECC under contracts No. SC1-CT91-0729 and No. SC1-CT92-0789.}

\end{Large}

\vspace{36pt}

\begin{large}
\renewcommand{\thefootnote}{\fnsymbol{footnote}}
H. Fujisaki, K. Kumekawa, M. Yamaguchi$^{\dag }$ \footnote[3]
{On leave of absence from
Department of Physics, Tohoku University},
and M. Yoshimura \\

Department of Physics, Tohoku University\\
Sendai 980-77 Japan\\
$^{\dag }$
Institut f\"{u}r Theoretische Physik\\
Physik Department, Technische Universit\"{a}t
M\"{u}nchen \\
D-85747 Garching Germany
\end{large}

\vspace{54pt}

{\bf ABSTRACT}
\end{center}

\vspace{0.5cm}
Thermal history after inflation is studied
in a chaotic inflation model with supersymmetric couplings of the inflaton
to matter fields.
Time evolution equation is solved in a formalism that incorporates
both the back reaction of particle production and the cosmological
expansion.
The effect of the parametric resonance gives rise to a rapid initial
phase of the inflaton decay followed by a slow stage of the Born
term decay.
Thermalization takes place immediately after the first explosive
stage for a medium strength of the coupling among created particles.
As an application we calculate time evolution of the gravitino abundance
that is produced by ordinary particles directly
created from the inflaton decay,
which typically results in much more enhanced yield than what a naive
estimate based on the Born term would suggest.

\newpage
{\bf \lromn 1. Introduction}

\vspace{0.5cm}
Inflationary universe models have several attractive features
\cite{inflation}:
it serves to resolve some outstanding problems
in the standard big bang model; how the homogeneous and isotropic
universe may emerge from a clumpy spacetime.
But how the inflationary stage is terminated and smoothly joined
to the hot big bang is not fully understood at present.
In particular, almost
all the entropy in our observable part of the universe is believed to be
created by dissipative oscillation of the inflaton field
around the minimum of a potential after the rapid cosmic expansion.
It is thus imperative to deepen our
understanding of the dissipative process of inflaton decay
in order to know more about the matter content of our universe.

The theory of particle production and associated field dissipation
has been much advanced recently \cite{linde et al 94},
\cite{brandenberger et al}, \cite{holman 95}, \cite{mine95-1},
\cite{reheating parametric}
by proper recognition of importance
of the phenomenon of the parametric resonance.
According to the new picture, one must
distinguish the two processes after inflation;
rapid particle production caused by
oscillation of the coherent inflaton field and subsequent thermalization
process. The thermalization occurs due to interaction and
decay of produced particles, which involves some weak coupling constant.
In contrast, an enormously large number of particles is created prior to
particle interaction by the
oscillating field due to the instability in the parametric resonance region,
if a combination of the initial oscillation amplitude of the inflaton and
its coupling to bose matter fields exceeds a critical value.
The time scale of resonant particle production and associated inflaton decay
is typically of the order of the inverse inflaton mass, which is short.
Thus the old picture \cite{reheat-original}
based on the Born term decay of inflaton must be modified.

In this paper we investigate thermal history after inflation and
demonstrate that Planck scale remnants such as the gravitino,
having rates ordinarily suppressed by the Planck mass, may be
overproduced from those directly produced by the inflaton decay.
It has been argued \cite{gravitino problem} for some time
that if inflation ends with a low enough reheating
temperature, the gravitino overproduction may be avoided in order not
to upset the success of the standard big bang cosmology.
We compute here the gravitino abundance right after inflation, starting
from the first principle of particle physics. It is found
that in a wide range of parameters
copious production of ordinary particles in an early epoch of
the inflaton decay gives rise to a significant amount of gravitino
abundance and subsequent dilution at later epochs is not very effective.
Our result on the gravitino abundance may equally be applied to any exotic
particles that have coupling to ordinary matter with the gravitational
strength.
Thus one may entertain a possibility that stable
secondary remnants created by direct products of the inflaton decay,
being capable of probing the Planck scale
physics, may actually comprise the bulk of the present mass density of the
universe as the non-baryonic dark matter.
If the gravitino is the Lightest Supersymmetric Particle (LSP), there
is a window in the parameter space for which the gravitino dominated
universe gives the solution to the nagging gravitino problem.

Although the complete theory of reheating still does not exist, it is necessary
to extend previous works on the inflaton decay and associated particle
production in order to discuss relic abundance of the Planck scale remnant.
A fundamental picture of the new theory of reheating was outlined
by Kofman, Linde, and Starobinsky \cite{linde et al 94}.
In our previous papers \cite{mine95-1}, \cite{fkyy95-1}
we discussed in great detail a field theoretical
framework of the inflaton decay and worked out time evolution, taking
into account the back reaction of produced particles.
In the following we first explain need to study in more detail both
the quartic and the Yukawa coupling cases in a class of supersymmetric models.
We then give our own account of the inflaton decay,
highlighted by a semi-classical set of differential
equations for the inflaton energy and the created radiation energy density
in the expanding universe.

Our present work, when limited to the inflaton decay, improves
over previous works mathematical
treatment of the parametric resonance in the large amplitude limit
and incorporates the small amplitude region at the same time.
Moreover, by integrating over short time behavior we derive tractable
time evolution equation for the inflaton and the created radiation
energy density, from which some physical consequences of the inflaton
decay, otherwise \cite{holman 95} difficult to derive, can be extracted.
Both the back reaction and the cosmological expansion play crucial
roles in our analysis.
Physical consequences that come out of our analysis are not much different
at the qualitative level from those of Ref \cite{linde et al 94}
in the first stage of inflaton decay,
but differ substantially in later stages.
Most notably, we find unlike others
that thermalization takes place immediately after
the first explosive stage of the inflaton decay, both because
in our models created particles interact among themselves
with ordinary strength of the medium magnitude,
and also because tremendous increase of the
particle number density occurs abruptly at a time in the first stage.

Relic gravitino abundance from inflation measured by the ratio of
the number densities,
\( \:
n_{3/2}/n_{\gamma }
\: \)
is roughly given by
\( \:
\sqrt{\frac{n_{\gamma }}{T}}/m_{{\rm pl}}
\: \)
with $n_{\gamma }$ the number density of produced massless particles and
$T$ a typical particle energy. This ratio agrees with the usual formula
\cite{gravitino problem}
\( \:
T_{R}/m_{{\rm pl}} \,,
\: \)
valid when thermalization is assumed with a reheat temperature $T_{R}$.
In the present work we first check how and when thermalization condition is
met and then follow time evolution of the gravitino abundance after
thermalization takes place. Fortunately the maximal particle production
and thermalization occurs almost at the same time in an early epoch of
the inflaton decay for a wide range of parameters, which makes our
analysis easier.
By solving time evolution equation,
one does not have to separate epochs of gravitino production
and its subsequent dilution, sometimes difficult to distinguish.
It turns out that the gravitino abundance thus computed cannot be
given by a simple formula like $T_{R}/m_{{\rm pl}}$.

It would be easier to understand the rest of this paper if we
first sketch essence of time evolution that comes out of our
subsequent detailed analysis. There are three important stages of
the inflaton decay.
After an initial period of the inflaton dominance,  abrupt decrease
of the inflaton amplitude occurs,
accompanying an enormous and the maximal rate of particle production.
Due to a high rate of interaction among produced particles thermalization
takes place immediately after this time.
The second stage is a slow process of particle production, that continues
until the last stage of complete decay when the Hubble time becomes
comparable to the inflaton decay time given by the Born formula.
Difference in the first two stages lies in the relevant $\theta $  value
(precisely defined later by
a dimensionless coupling in the mode equation of quantum field to be
created), large $\theta $ in the first stage of rapid decay and
small $\theta $ in the second stage of slow decay.
Although the bulk of the universe may be taken thermalized after the first
explosive stage, it would be misleading to assign a single reheat
temperature. Later we shall define two temperatures, the initial
reheat temperature $T_{i}$ immediately after the the explosive stage,
and the final reheat temperature $T_{f}$ after the complete Born decay.
For a large enough coupling the Planck scale relic is most abundantly
produced in the aftermath of the first explosive decay,
followed in the second stage
by a slow dilution of the abundance until the Born decay.
But this picture is modifed both for small coupling and for extremely
large coupling.

\newpage
{\bf \lromn 2. Model}

\vspace{0.5cm}
Let $\xi (t)$ a generic inflaton field treated here as a classical and
homogeneous variable.
Coupling of the inflaton to a generic bose matter field $\varphi $ is assumed
to contain both the Yukawa ($\xi \varphi ^{2}$) and the
quartic ($\xi ^{2}\varphi ^{2}$) type interaction, as so happens in
supersymmetric models we consider below.
Under a periodic $\xi $ oscillation the quantum state of $\varphi $
system exhibits instability due to the presence of the exponentially growing
mode in the parametric resonance region.
Evolution of the quantum ground state is simple
\cite{mine95-1}, \cite{fkyy95-1}:
the quantum state is described
by a density matrix that has rapidly oscillating off-diagonal elements.
Hence the averaged density matrix smoothed over short time scale
takes a nearly diagonal form.
Thus the system is almost classical: even a small bit of interaction
among created particles is expected to break the coherence and
yield the classical behavior of a mixed state.
Instability of the quantum $\varphi $ system is thus interpreted to give
rise to particle production and associated inflaton decay.
This occurs in a large region of the parameter
space of the coupling constant and the initial $\xi $ oscillation amplitude.
We use these results of particle production from the ground state even to
the case from mixed states such as the thermal state at late stages.
The basic attitude we take here is that a real process of particle production
may be insensitive to the state in which it occurs, while quantum fluctuation
involving virtual states of all kinds may sensitively depend on the state under
consideration.

For definiteness we concentrate in this work on the simplest kind of
inflation model, the chaotic inflation \cite{chaotic inflation}
with a mass term potential
\( \:
\frac{1}{2}\, m_{\xi }^{2}\,\xi ^{2}.
\: \)
It would be instructive to give beforehand
a crude range of parameters we have in mind. The fluctuation amplitude
in the density contrast observed by COBE suggests the inflaton mass
$m_{\xi }$ in the range of $10^{13}$ GeV \cite{cobe vs inflation},
while the initial oscillation
amplitude $\xi _{0}$ is of order the Planck scale.
Whenever we need concrete numbers, we use
\( \:
\xi _{0} = \sqrt{\frac{3}{4\pi }}\,m_{{\rm pl}} \approx  6\times 10^{18}
{\rm GeV}
\: \)
and $\xi _{0}/m_{\xi } = 5\times 10^{5}$.
The quartic coupling constant
$g^{2} $ and a dimensionless combination of the amplitude
\( \:
\theta_{0}  \approx g^{2} \xi _{0}^{2}/m_{\xi }^{2}
\: \)
are the key parameters for subsequent evolution after inflation. For
\( \:
g^{2} \gg  m_{\xi }^{2}/\xi _{0}^{2} \approx 10^{-12}
\: \)
the large amplitude oscillation with $\theta _{0} \gg 1$
is realized, thus our analysis in the following is applicable
in a large region of coupling constant.

A class of supersymmetric models we consider is characterized by
the superpotential of a generic form,
\( \:
W = \frac{1}{2}\, m_{\xi }\xi ^{2} + \frac{1}{\sqrt{2}}\,g\xi\varphi ^{2} \,,
\: \)
with $\varphi $ representing a chiral multiplet and $\xi $
the inflaton field. The magnitude of the coupling constant $g$ is not much
constrained and can be very small.
This gives ordinary potential terms of the form,
\( \:
\frac{1}{2}\, g^{2}\,\xi ^{2}\varphi ^{2} + \frac{1}{2}\,
gm_{\xi }\,\xi \varphi ^{2} \,,
\: \)
for the $\xi $ coupling.
Note that both the Yukawa and the quartic coupling terms are here with
a definite relation between the two.
The number of real scalar fields that couple to the inflaton field $\xi $
is taken to be 8 (the number of real components of the two Higgs doublets)
as in the minimal supersymmetric model.
Although we do not write interaction terms of bose fields $\varphi $
among themselves,
they are also induced by other terms of the Lagrangian besides the one
derived from this superpotential.
We assume that these generic bose fields have interaction strength of medium
size, $g_{s}^{2}/(4\pi ) = 1 - 10^{-2}$.
These interactions are important to establish rapid thermalization
once copious particle production occurs.
A quantity to measure  strength of the resonant particle production
is the variable mass term in the mode equation for the $\varphi $ field.
Ratio of the variable mass terms derived from
the quartic and the Yukawa coupling is
$ \approx g\xi /m_{\xi }$, which is large in the initial stage of the inflaton
decay.
Thus with an initial amplitude of
\( \:
g\xi _{0}/m_{\xi } \approx g\cdot 10^{6} \gg 1 \,,
\: \)
it is sufficient to consider the quartic coupling alone initially, although
the Yukawa coupling is important in later stages and ultimately
responsible for the Born term decay of
\( \:
\xi \:\rightarrow  \: \varphi \,\varphi \,.
\: \)
It would be worthwhile to recall the often cited reheat temperature of
\( \:
O[\,0.1\,]\,\sqrt{\,m_{{\rm pl}}\Gamma _{\xi }\,} \approx g\sqrt{N_{d}}
\cdot 10^{14}\,
{\rm GeV}
\: \)
with $N_{d}$ a number of decay channels, in the chaotic inflation model with
\( \:
m_{\xi } = 10^{13}\,{\rm GeV} \,.
\: \)
The problem with this naive formula is that it is not clear at what
stage of the inflaton decay
one should apply this reheat temperature, especially because
the new theory of inflaton decay implies a prolonged period of
particle production lasting until the complete Born decay.

\vspace{1cm}
{\bf \lromn 3. Parametric Resonance and Particle Production}

\vspace{0.5cm}
Analysis of the instability is divided into the small $\theta $
and the large $\theta $ regions, basic formulation of which
was respectively given in Ref \cite{mine95-1} and in Ref \cite{fkyy95-1}.
With the effective coupling term
\( \:
2g^{2}\xi ^{2}\varphi ^{2}
\: \)
specified in the initial stage,
the Fourier $\varphi $ mode of the wave vector $\vec{k}$ obeys
\( \:
\stackrel{..}{\varphi} _{k} + (\vec{k}^{2} +
4g^{2}\xi ^{2})\varphi _{k} = 0 \,,
\: \)
aside from effects of the cosmological expansion.
The effective $\xi ^{2}\varphi ^{2}$ vertex is different from the bare
vertex derivable from the bare Lagrangian, due to additional two diagrams
that contribute in the low energy limit with equal strength to the
bare vertex.
Taking the sinusoidal oscillation of the $\xi $ inflaton field
into account leads to the
standard form of Mathieu type equation,
\begin{eqnarray}
&&
\frac{d^{2}u}{d\tau ^{2}} + \left( h - 2\theta \cos (2\tau ) \right)
\,u = 0 \,, \\
&&
h = \frac{\omega ^{2}}{m_{\xi }^{2}} + \frac{2g^{2} \xi^{2}}
{m_{\xi }^{2}}
\,, \hspace{0.5cm}
\theta = \frac{g^{2} \xi ^{2}}{m_{\xi }^{2}} \,, \hspace{0.5cm}
\tau = m_{\xi }t \,,
\end{eqnarray}
with $\xi $ the amplitude of oscillation.
The relevant parameter region is thus restricted to
\( \:
h - 2\theta > 0
\: \)
for the quartic coupling.
In contrast the Yukawa coupling term becomes dominant in the late stages
and gives
\begin{eqnarray}
h = \frac{4\omega ^{2}}{m_{\xi }^{2}} \,, \hspace{0.5cm}
\theta = \frac{2g\xi }{m_{\xi }} \,, \hspace{0.5cm}
\tau = \frac{1}{2}\, m_{\xi }t \,,
\end{eqnarray}
thus the relevant region is not restricted to $h - 2\theta > 0 $.
The parametric resonance occurs in an
infinite number of bands of the instability
region separated by stability bands.
In the instability band the mode function exhibits the exponential
growth, $e^{\lambda \tau }$ with time $\tau $, and accordingly $\varphi $
particle production proceeds.
Thus important quantities one must extract from mathematical analysis of the
mode equation  are the growth rate $\lambda $ and the phase space
volume that contributes to the resonant $\xi $ decay.
The rate $\lambda $ in general depends on the energy $\omega $ of created
particle, but for a global analysis we perform in this work,
it is sufficient in some cases to know
a value of $\lambda $ averaged over the relevant phase space.

As was shown elsewhere \cite{fkyy95-1},
in the large amplitude limit of $\theta \gg 1$
one can replace the sinusoidal function in the potential term
of the Mathieu equation by a periodic potential, piecewise quadratic.
The exponential growth rate $\lambda $ of $\varphi $ particle production in the
resonance region is then given by
\begin{eqnarray}
&&
\lambda = \frac{1}{\pi }\,\ln (\,\sqrt{x} + \sqrt{x-1}\,) \,, \hspace{0.5cm}
x = (1 + e^{-\,\pi \sqrt{\theta }\,\epsilon })\,\cos ^{2}\psi \,,
\label{growth rate}
\\
&&
\psi = \frac{\pi ^{2}}{2}\sqrt{\theta } + \sqrt{\theta }\,\epsilon
\ln (\pi\,\theta ^{1/4}) +
\frac{1}{2}\, \Im \,\ln \left[ \,
\frac{\Gamma \left( \,(1 - i\sqrt{\theta }\epsilon )/2\,\right)}
{\Gamma \left( \,(1 + i\sqrt{\theta }\epsilon )/2\,\right)}\,\right] \,,
\end{eqnarray}
with
\( \:
\epsilon = \frac{h}{2\theta } - 1 \,.
\: \)
Even for a $\theta $ as small as $O[1]$ the formula above is a
qualitatively good approximation to the growth rate.
The parametric resonance region is determined by $x > 1$, which is
separated by the stability gap of $x<1$.
For instance, one may plot the growth rate $\lambda $ given by Eq
(\ref{growth rate}) against
\( \:
\epsilon
\: \)
for a fixed $\theta $, as illustrated in Fig.1.
Behavior of the growth rate $\lambda $ is very much different,
depending on whether $h > 2\theta  $ or $h < 2\theta $, in other words,
$\epsilon > 0$ or $\epsilon < 0$.
The relevant coupling in the large $\theta $ region in our supersymmetric
model is the quartic coupling,
which necessarily has $\epsilon > 0$, hence we shall concentrate on this
region here.
As a positive $\epsilon $ increases,
the peak growth rate $\lambda_{{\rm max}} $
within an island of the instability region
decreases, but the first peak $\lambda $ value always remains of order $0.2$.
This is related to that at $\epsilon = 0$, or along $h = 2\theta $,
the maximum value analytically calculated is given by
\( \:
\lambda _{{\rm max}} = \ln (\sqrt{2} + 1)/\pi \sim 0.28
\,.
\: \)
We numerically computed the average value of the growth rate in the
$\epsilon > 0$ region, taking a wide range of $\theta $, $10 - 10^{6}$.
The average of the first $\lambda $ peak value is $\approx 0.15$.
On the other hand, the phase space volume element multiplied by energy
to get the $\varphi $ energy density is
\( \:
2g^{4}\xi ^{4}/\pi ^{2}
\: \)
times $(0.07 - 0.1)/\theta $ numerically.

With cosmological evolution the $\theta $ value varies, and at late stages of
the inflaton decay one must consider the small $\theta $ region, even
if it starts in the large $\theta $ region initially.
Actually there are two causes of $\theta $
variation: either by decrease of the oscillation amplitude $\xi $
or by the $\xi $ mass ($m_{\xi }$) variation, which we later turn to.

The most important resonance band in the small $\theta $ region,
and the only one we consider in this paper is
the first band which is approximately given, with $\theta \leq 1$, by
\cite{mine95-1}
\begin{eqnarray}
{\rm Max}[\,0\,, h_{-}\,] < h
< h_{+} \,, \hspace{0.5cm}
h_{\pm } = 1 \pm  \theta - \frac{\theta ^{2}}{8} \,,
\end{eqnarray}
with $h$ and $\theta $ related to physical quantities of the Yukawa
coupling.
The phase space of the first band
as measured by the $h\, (\:\propto  \: \omega ^{2})$ width
becomes nonvanishing at
$\theta \approx \sqrt{24} - 4 \sim 0.90$, diminished at
\( \:
\sqrt{152} - 12 \sim 0.33 \,,
\: \)
until it becomes very small like $2\theta $ for very small $\theta $.

There is an important limitation in applying the resonant decay when
cosmological expansion is taken into account:
the cosmological redshift may push relevant frequency out of the
resonance band \cite{mine95-1}, \cite{brandenberger et al}.
The effect is particularly important in the second stage of slow decay.
The growth factor is roughly modified as follows.
Time interval during which a particular mode stays within the first band
at the Hubble time $1/H$ is given by
\( \:
\Delta t = \epsilon /(2H) = \theta /H \,,
\: \)
using the width of the first band $\epsilon $.
Multiplied by time averaged rate
\( \:
\lambda \approx \theta /4 \,,
\: \)
this gives the growth factor in the exponent,
\( \:
\approx m_{\xi }\theta ^{2}/(8H) \,.
\: \)
More precise estimate is possible by integrating the growth rate while
remaining within the narrow band during the frequency change
$\Delta \omega \sim \omega H\,\Delta t$,
\begin{eqnarray}
\frac{m_{\xi }}{2}\,\int\,dt\,\lambda = \frac{m_{\xi }}{4}\,
\int\,d\Delta \omega \,\frac{\sqrt{(h_{+} - h)
(h - h_{-})}}{|\frac{d\Delta \omega }{dt}|} =
\frac{\pi }{16}\,\frac{m_{\xi }\theta ^{2}}{H} \,,
\end{eqnarray}
where
\( \:
\Delta \omega = \omega - \omega _{{\rm res}}
\: \)
with $\omega _{{\rm res}} = m_{\xi }/2$ the center position of the
resonance.
The relevant phase space is estimated similarly.
For instance, it contributes to the created energy density,
\begin{eqnarray}
\frac{1}{2\pi ^{2}}\,\omega _{{\rm res}}^{3}\,|\frac{d\Delta \omega }{dt}|
= \frac{m_{\xi }^{4}}{32\pi ^{2}}\,H \,.
\end{eqnarray}
Thus with Ref \cite{brandenberger et al}
we may use for the source term of radiation,
\begin{eqnarray}
\frac{m_{\xi }^{4}}{32\pi ^{2}}\,H\,\sinh ^{2}[\,\frac{\pi }{16}\,
\frac{m_{\xi }\theta ^{2}}{H}\,] \,.
\label{small amp rate}
\end{eqnarray}

There is a problem with using this formula all the way down to the
smallest value of the amplitude $\xi $.
In the small $\theta = 2\sqrt{2}\,g\sqrt{n_{\xi }}/m_{\xi }^{3/2}$ limit
the number density $n_{\xi }$ of $\xi $ particles is very small, and
the coherence effect is not important.
One then expects that the Born rate for a single bosonic decay channel,
\( \:
\Gamma _{\xi }\,\rho _{\xi } = g^{2}m_{\xi }^{3}\xi ^{2}/(64\pi ) \,,
\: \)
should appear as the source term, leading to the ordinary
exponential decay law.
Indeed, it was shown \cite{mine95-1}
in the case without cosmological expansion taken into
account that the exponential decay law follows.
The source term above however gives a factor $\:\propto  \: g^{4}$ in the small
$g$ limit, which cannot be reconciled with the Born rate $\:\propto  \:
g^{2}$.
We suspect that a more refined treatment of the mode sum for the created
particle number in the expanding universe gives both the parametric resonance
effect in the medium small amplitude region and the Born rate behavior
in the extreme limit of small amplitude.
We have not succeeded in deriving this from a single formula
in a convincing way.
In the present work we are content to take a more phenomenological point
and simply add the
two terms, the formula above Eq (\ref{small amp rate})
and the Born rate, anticipating that
the two relevant regions do not overlap in the parameter space and
when one effect is large, the other is negligible.
This should be a practically good approximation to physical processes
involved, although a rigorous derivation is welcome.

Dissipation due to particle production necessarily
accompanies quantum fluctuation of the $\varphi $ field in the state.
As noted in Ref \cite{linde et al 94}, \cite{holman 95},
this has an important effect on variation of the $\xi $ mass via
\( \:
g^{2}\langle \varphi ^{2} \rangle
\: \).
The expectation value here may be evaluated by using an exact relation
in each mode between the produced particle number and the fluctuation,
which follows
since the wave function of the quantum state is given by a Gaussian form.
With the exponentially small term neglected, this relation reads as
\( \:
\langle N_{\omega } \rangle \sim \frac{\omega ^{2} + \lambda ^{2}m_{\xi }^{2}}
{2\omega }\,\langle  q_{\omega }^{2} \rangle \,,
\: \)
with $q_{\omega }$ the coordinate variable of harmonic oscillator of
mode $\omega $.
In both the large $\theta $ or the small $\theta $ region one can
neglect $\lambda $ dependent term here.
First, in the large amplitude region typical resonance energy
\( \:
\omega ^{2} = O[g\xi m_{\xi }] \gg m_{\xi }^{2} \,.
\: \)
Second, in the small amplitude limit
\( \:
\lambda m_{\xi } \approx  \theta m_{\xi }/4 \,,
\: \)
which becomes smaller than $m_{\xi }$ itself.
In the following we denote the variable mass by $m_{\xi }\mu $.
The mass variation in the region where the Yukawa coupling dominates
involves one loop two point function in which two external $\xi $ lines
are separated unlike the previous tadpole term.
Computation of this particular diagram in the quantum $\varphi $ state
in question has not been worked out.
We replace the entire one loop contribution by the single tadpole term
$\langle \varphi ^{2} \rangle$ we calculate here,
hoping that the neglected one loop term gives a contribution
comparable to, and not too much larger than, the calculated tadpole.
If this is the case, our neglect of the other contribution would be
excused, because the mass variation has been found not to be drastic,
$\mu < O[100]$ in the coupling region of our interest.
Even more precise mass variation term would not change much of our conclusion.

Summarizing the particle production rate of $N_{p}$ species
($= 8$ in the minimal supersymmetric model)
and  the rate of mass variation,
we have obtained the mode-summed expression:
with dimensionless functions defined by
\begin{eqnarray}
&&
y = \rho _{\xi }/\rho _{\xi }^{0} \,, \hspace{0.5cm}
z = \rho _{r}/\rho _{\xi }^{0} \,, \hspace{0.5cm}
\rho _{\xi }^{0} = \frac{1}{2}\, m_{\xi }^{2}\xi _{0}^{2} \,, \hspace{0.5cm}
\mu = m_{\xi }(t)/m_{\xi } \,, \\
&&
\frac{d}{d\tau }\,\sum_{\varphi }\,\langle \rho _{\varphi } \rangle
\equiv
\frac{1}{2}\, m_{\xi }^{2}\xi _{0}^{2}\,B\,s_{y}
\,, \hspace{0.5cm}
g^{2}\,\frac{d}{d\tau }\,\sum_{\varphi }\,\langle \varphi ^{2} \rangle
\equiv
2\,m_{\xi }^{2}\,A\,s_{\mu } \,,
\end{eqnarray}
in the large $\theta $ region
\begin{eqnarray}
&&
A
\approx \frac{0.17 \times 0.3}{\pi ^{2}}\,2N_{p}\,
g^{3}\,\frac{\xi _{0}}{m_{\xi }}
\,, \hspace{0.5cm}
B
\approx \frac{0.1 \times 0.3}{\pi ^{2}}\,4N_{p}\,g^{2} \,, \\
&&
s _{y} = y\mu \,\exp [\,0.3 \,\int_{0}^{\tau }\,d\tau'\, \mu (\tau ')\,]
\,, \hspace{0.5cm}
s_{\mu } = \sqrt{y}\,\exp [\,0.3 \,\int_{0}^{\tau }\,d\tau'\, \mu (\tau ')\,]
\,,
\end{eqnarray}
by taking an average value 0.3 for $2\lambda $,
while in the small $\theta $ region
\begin{eqnarray}
&& \hspace*{1cm}
A \sim \frac{N_{p}\,g^{2}}{64\pi ^{2}} \,, \hspace{0.5cm}
B \sim \frac{N_{p}}{256\pi ^{2}}\,(\frac{m_{\xi }}{\xi _{0}})^{2}
\,, \\
&& \hspace*{-1.5cm}
s_{y} = \mu ^{4}\sqrt{y+z}\,
\sinh ^{2}[\pi \theta _{0}\,\frac{y}{\mu ^{5}\sqrt{y+z}}\,]
\,, \hspace{0.5cm}
s_{\mu } = \mu \sqrt{y+z}\,
\sinh ^{2}[\pi \theta _{0}\,\frac{y}{\mu ^{5}\sqrt{y+z}}\,] \,,
\end{eqnarray}
with understanding that the Born term should be added in the source
term of evolution equation.

\newpage
{\bf \lromn 4. Inflaton Decay and Thermal History after Inflation}

\vspace{0.5cm}
Time evolution of the inflaton and the radiation energy density
is given by a coupled set of differential equations.
This equation is derived by considering energy balance between $\xi $
and $\varphi $ with the cosmological expansion taken into account.
Variation of the $\xi $ mass involves the expectation value
$\langle \varphi ^{2} \rangle$
which varies
like
\( \:
a^{-2}(\tau ) \,,
\: \)
as the cosmological scale factor $a(\tau )$ increases.
Coupled evolution of the $\xi $ energy, the $\varphi $ energy density, and the
$\xi $ mass is then described in terms of the dimensionless quantities by
\begin{eqnarray}
&&
y' + 3d\sqrt{y+z}\,y = - Bs_{y} - d\gamma \,y \,,
\\
&&
z' + 4d\sqrt{y+z}\,z = Bs_{y}  + d\gamma \,y \,,
\\
&&
\mu' + d\sqrt{y+z}\,(\mu - \frac{1}{\mu }) = As_{\mu } \,,
\end{eqnarray}
with
\( \:
d \equiv \sqrt{\frac{4\pi }{3}}\,\frac{\xi _{0}}{m_{{\rm pl}}} \,,
\: \)
and
\( \:
\gamma = 2N_{p}g^{2}/(32\pi )
\: \)
the dimensionless Born decay rate.
The number of $\xi $ decay channels is doubled here since fermions
of supersymmetric partners contribute to the Born rate, but not to
the parametric resonance effect.
The initial time for integration may be set at the Hubble $H = m_{\xi }$,
since the damped oscillation effectively starts there.
This amounts to letting $d = 1$, assuming no initial radiation present;
\( \:
y(0) = 1 \,, z(0) = 0
\: \).

The matching from the large to the small amplitude region is somewhat
delicate. Recall the definition of $\theta $ for the Yukawa and the quartic
couplings, which leads to a relation,
\( \:
\theta _{Y} = 2\,\sqrt{\theta _{4}}/\mu \,.
\: \)
Suppose that the mass variation is small, $\mu \sim 1$.
It is then possible to have a situation that despite of $\theta _{4} < 0.33$
the Yukawa $\theta _{Y} > 0.33$. We should ensure for the matching
that the larger one of $\theta _{Y}$ or $\theta _{4}$ is small, and
within the small amplitude region.
Without embarking on a too much elaborate analysis,
perhaps not worthy of its effort at
the present level of investigation, we practically adopt as the small amplitude
condition $\theta _{Y} < 0.33$ and use the large amplitude
formula for the quartic coupling.
For completeness we did also integrate the differential equation by
a more elaborate matching; in the intermediate region with
$\theta _{4} < 0.33 $ and $\theta_{Y} > 0.33$ we have replaced the source term
by formulas applicable in the large Yukawa coupling region, as worked out
in Ref \cite{fkyy95-1}. Difference in this case is appreciable only
in a narrow region of the coupling $g$, as seen in Fig 7b for the
final gravitino abundance.
Hence the final outcome does not seem to sensitively depend on
the matching condition.

We did some extensive numerical investigation of time evolution equation
for the set of parameters, $g$ and $\xi _{0}/m_{\xi }$.
Here we restrict to the special case of
$\xi _{0}/m_{\xi } = 5 \times 10^{5}$ and $g = O[\,1 - 10^{-5}\,]$,
consistent with COBE observation in the chaotic inflationary scenario.
With a coupling less than $10^{-5}$
the effect of the parametric resonance is not significant.
This critical strength for the parametric resonance may be determined
by the condition, $\theta _{4} \gg 1$, leading to
\( \:
g \gg m_{\xi }/\xi _{0} \approx 10^{-5} \,.
\: \)
Moreover, it has been found by the present work
that not only for $g<10^{-5}$ but also for $g>1$, the naive
estimate of the gravitino abundance by the Born formula happens to be
numerically close to the correct value.
This is the reason we concentrate on the coupling region of
\( \:
1 > g > 10^{-5} \,.
\: \)
Before we present result of numerical computation, we discuss question of
thermalization, coherence and how to compute the gravitino abundance.

Although the process towards complete thermalizaion is not fully
understood at present,
the following argument suggests that with $\rho _{r} \approx  \rho _{\xi }$
at the exit $t = t_{e}$ from the first explosive stage,
thermalization takes place immediately.
For a wide range of the coupling, the condition
$\rho _{r} \approx  \rho _{\xi }$ holds, for instance for
$g > 3\times 10^{-5}$, $10 > \rho _{r}/\rho _{\xi } > 0.01$.
Note first that a typical energy of created particles at the exit time,
not necessarily thermalized as yet, is of order
\( \:
T \approx \sqrt{gm_{\xi }\,\mu \xi } \approx \theta ^{1/4}\,m_{\xi }\,\mu
\approx m_{\xi }\,\mu \,.
\: \)
Interaction rate among $\varphi $ particles
compared with the Hubble rate is very crudely estimated as
\begin{eqnarray}
\frac{\Gamma }{H} \approx \frac{n_{r}/T^{2}}{\sqrt{\rho _{\xi } + \rho_{r} }
/m_{{\rm pl}}}
\approx \frac{\sqrt{\rho_{r} }\,m_{{\rm pl}}}{\sqrt{1+1/r}\,T^{3}}
\approx \frac{r}{\sqrt{1+r}}\,
\frac{\xi _{e}\,m_{{\rm pl}}}{m_{\xi }^{2}\,\mu _{e}^{2}}
\approx \frac{r}{\sqrt{1+r}}\,\frac{m_{{\rm pl}}}{g\,m_{\xi }\mu _{e}}
\,,
\end{eqnarray}
with $r = \rho _{r}/\rho _{\xi }$
the ratio of the radiation to the $\xi $ energy density at the exit.
Thus with $\xi _{0} $ close to the Planck scale,
\( \:
\Gamma \gg H \,
\: \)
at $t_{e}$ since
\( \:
\mu _{e} < O[10^{2}]
\: \)
in the parameter range of our interest, and thermalization takes place
almost at the same time as the explosive resonant particle production.
We did a consistency check of the condition of thermalization
\( \:
\Gamma /H \gg 1
\: \)
in our numerical analysis, by using the interaction
rate $\Gamma = \alpha T$ with
$\alpha > 10^{-2} $ and $T$ the temperature estimated
from the radiation energy density $\rho _{r}$.
It turns out that this thermalization condition is always obeyed
after the explosive exit from the large amplitude region, as shown in Fig 2.

Even the multi-particle reaction that works very efficiently as the
energy exchange process may occur frequently. For instance,
rate of the (N+1) $\:\rightarrow  \:$ 2 body reaction behaves
on dimensional grounds like
\begin{eqnarray}
\alpha _{s}T\,(\frac{\alpha _{s}N_{s}}{\pi }\,\frac{n}{T^{3}})^{N} \,,
\end{eqnarray}
with $\alpha _{s} = g_{s}^{2}/(4\pi )$ a typical medium size coupling
constant, $n$ the number density of $N_{s}$ species of colliding particles,
$T$ an average energy of colliding particles.
The point of this discussion is that in situations immediately after
the explosive stage, but before thermalization,
there are more low-energy particles than what is expected from
the thermal distribution, which implies that $n \gg T^{3}$.
With $\frac{\alpha _{s}N_{s}}{\pi }\,\frac{n}{T^{3}} \gg 1$,
one may expect that both the chemical and the thermal equilibrium
is established rapidly.
The number of reactions needed to reach equilibrium may crudely be
estimated as follows.
A typical amount of exchanged energy per reaction is very roughly
$T \approx m_{\xi }\mu _{e}$ as noted above. To get the thermal energy
of order $\rho _{r}^{1/4}$, one needs energy exchange reactions
$N_{r}$ times, with
\begin{equation}
N_{r} \sim \frac{\rho _{r}^{1/4}}{T} \approx \frac{\sqrt{m_{\xi }
\mu _{e}\xi _{e}}}{m_{\xi }\mu _{e}} \approx \frac{1}{\sqrt{g}} \,,
\end{equation}
under the assumption of $\rho _{r} \approx \rho _{\xi }$ at the exit.
This is not a huge number, at most 100, and one may expect that
equilibrium is reached fast.

Time evolution of the thermal temperature $T$, defined
from the radiation energy density by
\( \:
T = [\,\rho _{r}/(\frac{\pi ^{2}}{30}\,N)\,]^{1/4}
\: \)
with $N = 200$,
is shown for a few choices of the coupling in Fig 3, together with evolution of
the net entropy $sa^{3}(t)$; the true measure of dissipation.
In all these cases plateau regions of the net entropy are clearly observed
indicating period of quiet evolution, but behavior of the
transitional era is different depending on the coupling $g$.
Although the temperature is calculated from the radiation energy density
at all times, thermalization does not take place prior to the explosive
stage, hence the temperature in this early epoch has no meaning.

Despite of our detailed control of time evolution it would be useful
to present what is left in the end after the second
reheating due to the Born decay.
The final reheat temperature defined at a time when the net entropy
\( \:
\rho _{r}^{3/4}\,a^{3}
\: \)
settles down to its asymptotic value (effectively replaced by that at a time,
10$\times $ Born lifetime),
is thus plotted against the coupling $g$ in Fig 4.
Remarkably, this is not much different from
the usually estimated reheat temperature of order
\( \:
0.1\,\sqrt{\,m_{{\rm pl}}\,\Gamma _{\xi }\,} \,,
\: \)
using the Born formula.
We however warn that this does not necessarily
support the simple picture of the inflaton decay based on the Born term:
in the intermediate times between the explosive stage and the Born stage
there exits a prolonged epoch with higher temperatures.
Physical processes that involve higher temperatures such as baryo-genesis
or gravitino production must be studied taking into account this
period.
For reference we plotted in Fig 5 the initial reheat temperature $T_{i}$
defined immediately after the explosive stage.
The maximal temperature we observe in the whole range of the coupling $g$
is $\approx 90\,m_{\xi } \approx 10^{15}$ GeV close to the GUT scale,
right after the explosive stage.
As repeatedly stressed here and will be explicitly demonstrated later for
the gravitino abundance,
the correct thermal history cannot be described
in terms of a single temperature, either $T_{i}$ or $T_{f}$.

Despite of immediate thermalization after the explosive first stage
decay, energy balance equation derived without considering effect of
thermalization is presumably little affected since
the energy conservation holds globally irrespective of energy exchange
process towards thermalization.
On the other hand, the expectation value
\( \:
\langle \varphi ^{2} \rangle
\: \)
may differ considerably in thermal equilibrium from that we derived
previously, which may then modify the equation for the mass variation.
For simplicity we ignore this thermal effect in the present work.

Coherence of the cosmic $\xi $ field is crucial to the phenomenon of
the parametric resonance, especially in the large amplitude region.
One may estimate in the following way how and in what parameter region
the coherence is lost.
First, one may regard the homogeneous and coherent field as an aggregate
of massive particles at rest with a phase coherence precisely tuned.
Suppose that some of these $\xi $ particles interact with created $\varphi $
particles, which partially breaks the precise phase coherence.
Thus whenever interaction rate $\Gamma_{\xi } $
between $\xi $ and $\varphi $ particles
dominates over the cosmic expansion rate $H$, the coherence is
considered to be lost.
To estimate the interaction rate
\( \:
\Gamma_{\xi } = \langle \sigma v \rangle\,n_{\varphi }N
\: \)
with $N = 200 \,, $
one first computes the cross section $\sigma $.
As to the relevant energy of $\varphi $ particles the thermal average,
$\sim 2.7\,T$, may be taken,
which becomes as large as $\approx 100\,m_{\xi }$ in
some situations we consider.
Thus for simplicity
we take the high energy limit $\omega \gg m_{\xi }$ of the cross section.
This results in the cross section,
\( \:
\frac{g^{4}}{2\pi \,m_{\xi }\omega } \,.
\: \)
The $\xi $ interaction rate relative to the Hubble rate is then estimated as
\begin{eqnarray}
\frac{\Gamma_{\xi } }{H} = 0.68\,
\frac{\zeta (3)}{2\pi ^{4}}\,\sqrt{15N}
\,g^{4}\,\frac{\xi _{0}}{m_{\xi }} \approx 1.2\times 10^{5}\,g^{4}
\,,
\end{eqnarray}
assuming the radiation dominated thermal bath of temperature $T$.
Thus one has to worry about the loss of coherence only for a large
coupling of order $0.05$ or larger.
We have checked more carefully at what coupling $g$ the coherence is lost,
by using numerically computed Hubble rate.
In most of the parameter region except for $g > 0.08$
the ratio of the $\xi $ interaction rate to the Hubble rate does not
exceed unity.
In our presentation of various output quantities we ignore the coherence
constraint, since for these large couplings of $g > 0.08$
the gravitino abundance itself is too large even if
the parametric resonance effect is altogether neglected.

\vspace{1cm}
{\bf \lromn 5. Gravitino Abundance}

\vspace{0.5cm}
Time evolution of the gravitino abundance follows the rate equation,
\begin{eqnarray}
\frac{dn_{3/2}}{dt} + 3H\,n_{3/2} = \langle \Sigma v \rangle\,
n_{\varphi }^{2} \,,
\end{eqnarray}
since the annihilation term may be neglected due to the low rate
of production, as consistently checked by the outcome,
\( \:
n_{3/2} \ll n_{\varphi }
\: \).
The cross section $\Sigma  $ contains all contributing channels, and
it was computed as
\( \:
\Sigma = 250/m_{{\rm pl}}^{2}
\: \)
\cite{kawa-moroi}.
The remaining uncertainty of the $\varphi $ number density $n_{\varphi }$
is fixed by using the relation to the energy density in a
thermal bath, valid at times after the explosive stage;
\begin{eqnarray}
n_{\varphi } = \frac{\zeta (3)}{\pi ^{2}}\,(\frac{30\rho _{r}}
{\pi ^{2}N})^{3/4} \,.
\end{eqnarray}
We plotted in Fig 6 time evolution of the gravitino abundance for
a few choices of the coupling.
An unexpected result is that for large enough couplings a sizable
gravitino yield produced immediately after the explosive stage
is almost completely canceled by the late
Born decay, resulting in the final abundance nearly equal to the naive
rate estimated by the Born term.

The final gravitino abundance against the coupling is shown in Fig 7,
together with the abundance obtained without considering the resonance effect.
In other words, we also evaluate for comparison
time evolution of the gravitino abundance
with the Born decay term $\gamma _{B}\,\rho _{\xi }$
as the source, without assuming either the instantaneous
decay or the instantaneous reheating.
This way of computing the gravitino abundance would give more precise
prediction of the Born term decay than the usual simple estimate
based on the instantaneous decay and reheating.
The final gravitino abundance is substantially larger
in the range of couplings, $10^{-4} < g < 0.1$
than what the Born decay would suggest (approximately $8\times 10^{-8}\,g$).
It should be noted that gravitino production and its subsequent
dilution is a process that goes on continuously from the first explosive
stage to the final stage of the inflaton decay.
Although the final reheat temperature is not too much
different from the Born estimate, what happens in the intermediate
stages may differ for different couplings,
and the final abundance has a complicated behavior as a function
of the coupling.

Fate of the produced gravitinos has been much discussed in the literature.
There are strong constraints from the decay products of the photon
\cite{gravitino-photon const} and
LSP (Lightest Supersymmetirc Particle) \cite{gravitino-lsp}.
Production of both the electromagnetic and the quark showers
\cite{gravitino-quark const} around the epoch of
nucleosynthesis must be much suppressed in order not to destroy produced light
elements. The decay mode containing showers of strongly interacting particles
is very much model dependent due to the unknown scalar quark and the gluino
mass spectrum, while  photo-dissociation of D, $^{3}$He, $^{4}$He gives
a strong constraint on the abundance $n_{3/2}/s$ with $s$ the entropy
density if the gravitino mass is less than a few TeV \cite{kawa-moroi}.

Here we merely mention the constraint from a possible overclosure
\cite{gravitino-lsp}
of the present mass density of the universe due to the stable LSP
denoted here by $\chi $,
coming from the gravitino decay.
Recall first the decay rate, valid for the gravitino mass
$m_{3/2} \ll   M = m_{{\rm pl}}/\sqrt{8\pi }$,
\begin{eqnarray}
\Gamma (\psi _{\mu } \:\rightarrow  \: \chi  + A_{\nu }) =
\frac{1}{32\pi }\,\frac{m_{3/2}^{3}}{M^{2}} \sim
(4 \times 10^{8} \,{\rm sec})^{-1}\,(m_{3/2}/100\,{\rm GeV})^{3} \,,
\end{eqnarray}
where the specified decay mode is the gauge boson ($A_{\nu }$) and
the gaugino ($\chi  $).
Relic LSP abundance $n_{\chi }/s$ is given either by
the abundance of the parent gravitino $n_{3/2}/s$ if this yield is
not too large, or by the thermal annihilation abundance if the
initial abundance is large,
\begin{eqnarray}
&&
\frac{n_{\chi }}{s} = {\rm Min}\;
\left( \frac{n_{3/2}}{s}\,, (\frac{n_{\chi }}{s})_{{\rm ann}}\right)
\,, \hspace{0.5cm}
(\frac{n_{\chi }}{s})_{{\rm ann}} \sim 3.8\,g_{*}^{-1/2}\,\frac{1}
{\langle \sigma_{{\rm ann}} v \rangle\,m_{{\rm pl}}\,T_{D}} \,, \\
&&
\Omega _{\chi }h^{2} \sim 3 \times 10^{10}\,(\frac{m_{\chi }}{100{\rm GeV}})
\,\frac{n_{\chi }}{s} \,.
\end{eqnarray}
The annihilation abundance $(n_{\chi }/s)_{{\rm ann}}$ has been
estimated from the balance between the
annihilation and the Hubble rates at the freeze-out $T = T_{D}$,
\( \:
\langle \sigma _{{\rm ann}}\,v \rangle\,n_{\chi } = H(T_{D}) \,,
\: \)
with $g_{*} $ the number of massless species at the freeze-out.

Let us take as an example the LSP mass $m_{\chi } = 100\,$GeV.
{}From Fig 7 we find that the condition of non-overclosure with
\( \:
n_{\chi } = n_{3/2}
\: \)
implies $g < (1 - 2)\times 10^{-4}$.
If this bound is violated, one must reduce the LSP abundance via
thermal annihilation that may later occur, in order not to contradict
with the closure density.
For simplicity let us take a likely possibility of $\chi $ being the bino,
in which case the annihilation rate at temperature $T$ may be given by
\begin{eqnarray}
\langle \sigma_{{\rm ann}} v \rangle =
a + b\frac{T}{m_{\chi }} + O(\frac{T}{m_{\chi }})
^{2}
\,, \hspace{0.5cm}
b = \sum_{f}\,b_{f\bar{f}} \sim 10^{-2}\,\frac{m_{\chi }^{2}}{(m_{\chi }^{2}
+ m_{\tilde{f}}^{2})^{2}} \,,
\end{eqnarray}
assuming a common mass $m_{\tilde{f}}$ of exchanged particles.
The condition of non-overclosure then leads to
\begin{eqnarray}
m_{3/2} > 100\,{\rm TeV} \,.
\end{eqnarray}
We took over the result of a similar analysis of
Ref \cite{gravitino-lsp-recent}, applied for the Polonyi field decay.
Thus with $m_{\chi } = O[100\,$GeV], one must have either a very small coupling
$g < 10^{-4}$ or a gravitino mass larger than 100$\,$TeV, which
is not favored from the point of low energy SUSY breaking.
The constraint becomes even more stringent if one also considers destruction
of light elements by electromagnetic and hadronic showers due to the
gravitino decay.

More interesting is the possibility of the gravitino being LSP
\cite{gravitino-photon const}.
The old overclosure bound $m_{3/2} < 1\,{\rm keV}$
\cite{gravitino mass bound-old}
is based on the thermal abundance of gravitino production, hence is not
applied at the face value
if inflation dilutes away thermal gravitinos, as in our case.
Besides the primordial production immediately after inflation computed by us,
there may be a new source of the gravitino production due to the
next-to-the-lightest superparticle (NSP) decay. The production rate of
this process has been found negligible
and the corresponding bound on the gravitino mass, including argument
of photo-dissociation due to radiative decay component,
have been given in Ref \cite{stable gravitino}.
We should however derive the bound, not in terms of the reheat temperature
$T_{R}$ and the gravitino mass, but in terms of the coupling $g$ and
the mass, since the reheat temperature is not well defined in the new
theory of reheating.
The result is shown in Fig 8.
In this argument it is important \cite{stable gravitino}
to modify the gravitino production cross section to
\begin{equation}
4.5\,\frac{m_{\tilde{g}}^{2}}{m_{3/2}^{2}\,m_{{\rm pl}}^{2}} \,,
\end{equation}
assuming that gravitino production proceeds mainly from gluino, as
is the case unless the gluino mass $m_{\tilde{g}}$ is too small.
The reason the gravitino mass $m_{3/2}$ appears in the denominator is
that with a light gravitino the production mainly proceeds via
longitudinal component, Goldstino.
Although not shown in Fig 8, the closure condition
$\Omega _{3/2}h^{2} = 1$ is met roughly with
\( \:
g/(m_{3/2}/1\,{\rm GeV}) = 3\times 10^{-5}
\: \)
even for $m_{3/2} < 10^{-2}\,$GeV, down to $m_{3/2} \sim 1\,$keV at which
the thermal, hence the maximal yield is reached.
Thus with the gravitino mass $m_{3/2} < 1\,$keV, thermal production becomes
possible even after inflation.
Also is not shown the constraint from the photo-dissociation of light
elements which becomes very stringent for a gravitino mass less than
a few GeV \cite{kawa-moroi}.
The maximal reheat temperature right after the explosive stage can
become as large as $\approx 2\times 10^{14}\,$GeV, with $g \approx 6\times
10^{-5}
\,, m_{3/2} \approx 1\,$GeV (for $m_{\tilde{g}} = 150\,$GeV) in the
gravitino dominated universe. This high temperature might help to generate
the baryon asymmetry at the GUT scale.
The gravitino dominated universe thus
remains as an exciting solution to the nagging gravitino problem and
deserves further study.
We note that our calculation is presumably the first instance that
computed the gravitino abundance starting from the first principle.

\vspace{1cm}
In summary, the thermal history after inflation has been elucidated in a simple
model of chaotic inflation that contains a generic type of coupling to
matter fields.
The outcome of the inflation cannot be described by a single reheating
temperature, since the initial phase dominated by the parametric resonance
effect, ends with much higher temperatures
followed by a slow inflaton decay due to the usual Born term.
High energy processes are much affected in intermediate stages, and
we explicitly demonstrated this by following time evolution of the gravitino
abundance.


\newpage

\begin{Large}
\begin{center}
{\bf Figure caption}
\end{center}
\end{Large}

\vspace{0.5cm}
\hspace*{-0.5cm}
{\bf Fig 1}

Growth rate in the instability region plotted against
$\epsilon = \frac{h}{2\theta } - 1$ for a fixed value of
$\theta = 1000$.

\vspace{0.5cm}
\hspace*{-0.5cm}
{\bf Fig 2}

Thermalization condition measured by the quantity, the thermal temperature
relative to the Hubble rate
\( \:
T/H
\: \)
that closely approximates the interaction rate,
$\Gamma _{\varphi }/H$.

\vspace{0.5cm}
\hspace*{-0.5cm}
{\bf Fig 3}

Time evolution of the temperature in the unit of the inflaton mass $m_{\xi }$
(solid line) and the net entropy defined
by the entropy density times the scale factor to the 3rd power (dash-dotted)
for three cases of the coupling
(A) $g = 10^{-2}$, (B) $g = 10^{-3}$
(C) $g = 10^{-4}$.
The broken line shows the temperature as computed from the source term
of the Born decay without effect of the parametric resonance.
The time scale is given in the unit of
the Born decay lifetime $\frac{32\pi }{16\,g^{2}m_{\xi }}$.

\vspace{0.5cm}
\hspace*{-0.5cm}
{\bf Fig 4}

Final temperature in the unit of $m_{\xi }$ plotted against the coupling
$g$.
The final time is taken at 10 $\times \tau _{B}$ (Born lifetime).
Result based on the Born term decay depicted by the broken line,
coincides with the more precise result (solid line) within precision of
this figure.

\vspace{0.5cm}
\hspace*{-0.5cm}
{\bf Fig 5}

Initial temperature in the unit of $m_{\xi }$
immediately after the explosive stage, at the time of
\( \:
\theta _{Y} = 0.33
\: \).

\vspace{0.5cm}
\hspace*{-0.5cm}
{\bf Fig 6}

Time evolution of the gravitino abundance (solid line) plotted against time
the unit of the Born decay lifetime $\frac{32\pi }{16\,g^{2}m_{\xi }}$,
for three couplings
(A) $g = 10^{-2}$, (B) $g = 10^{-3}$
(C) $g = 10^{-4}$.
For comparison result due to the Born term alone is shown by the
broken lines.

\vspace{0.5cm}
\hspace*{-0.5cm}
{\bf Fig 7A}

Final gravitino abundance relative to the entropy density plotted
against the coupling $g$ (solid line).
The final time is taken at 10 $\times \tau _{B}$ (Born lifetime).
For comparison is shown the same abundance calculated without
the source term due to resonance effects (broken line).

\vspace{0.5cm}
\hspace*{-0.5cm}
{\bf Fig 7B}

The same gravitino abundance as in Fig 7B,
with the two step matching as described
in the text (solid line), to be compared with the one step matching in
Fig 7A (dash-dotted line), and with the Born rate (broken line).

\vspace{0.5cm}
\hspace*{-0.5cm}
{\bf Fig 8}

Region of the parameters $g \,, m_{3/2}$ in which the stable gravitino
overcloses the present mass density of universe (shaded region).
The solid lines indicate
the closure contours with the two choices of the gluino mass
$m_{\tilde{g}}$, and
the broken lines the similar ones obtained from the Born source term.

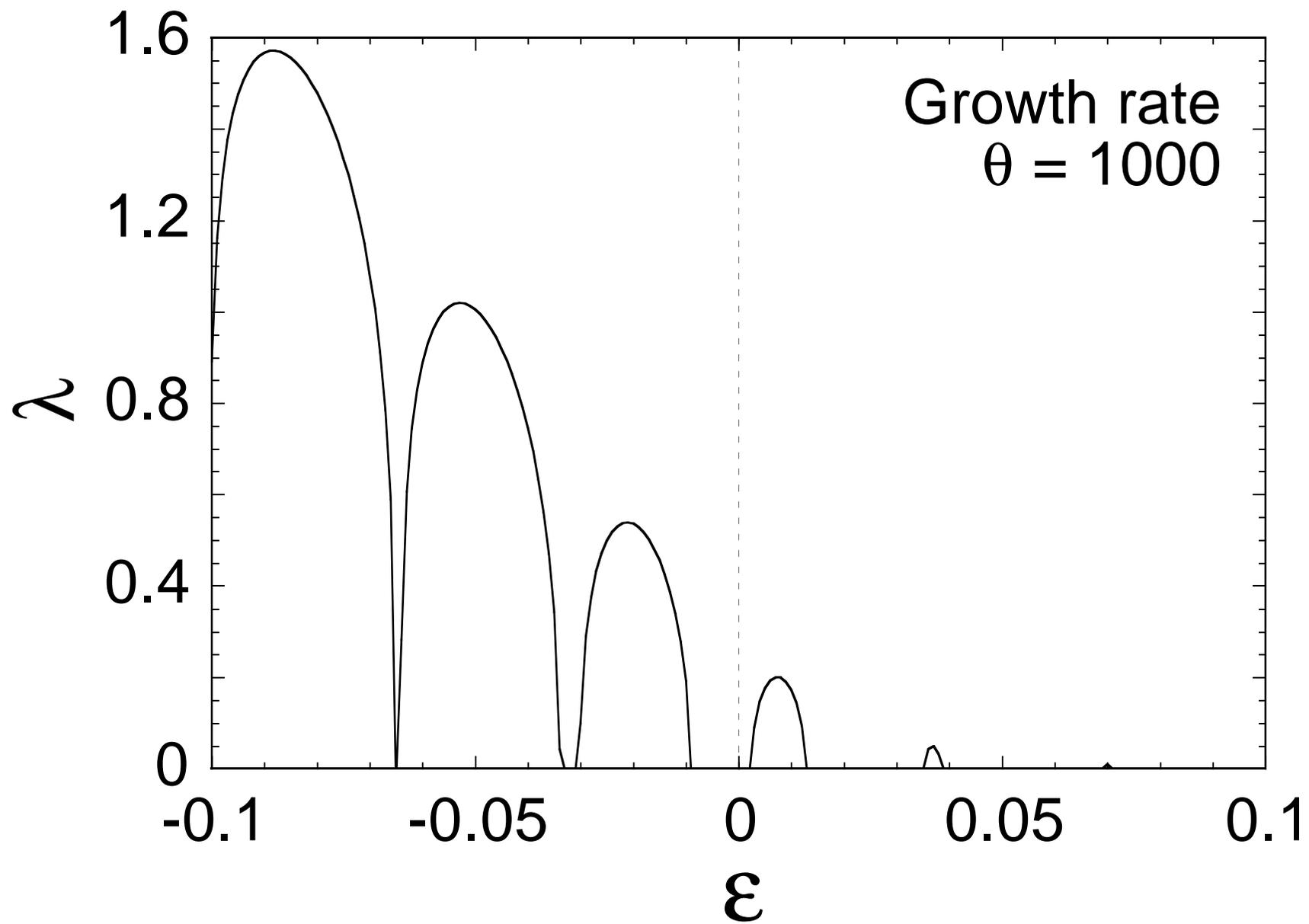
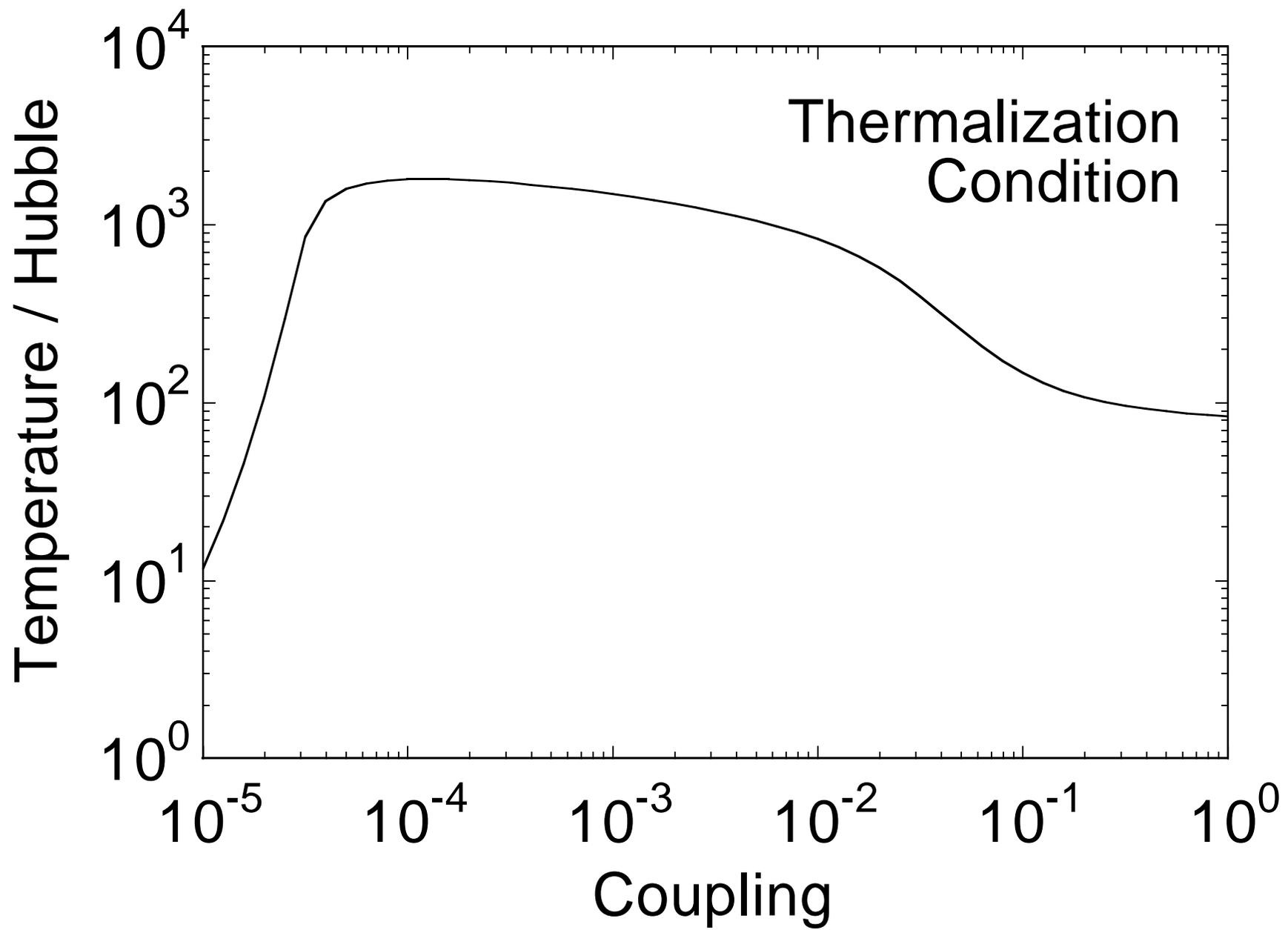
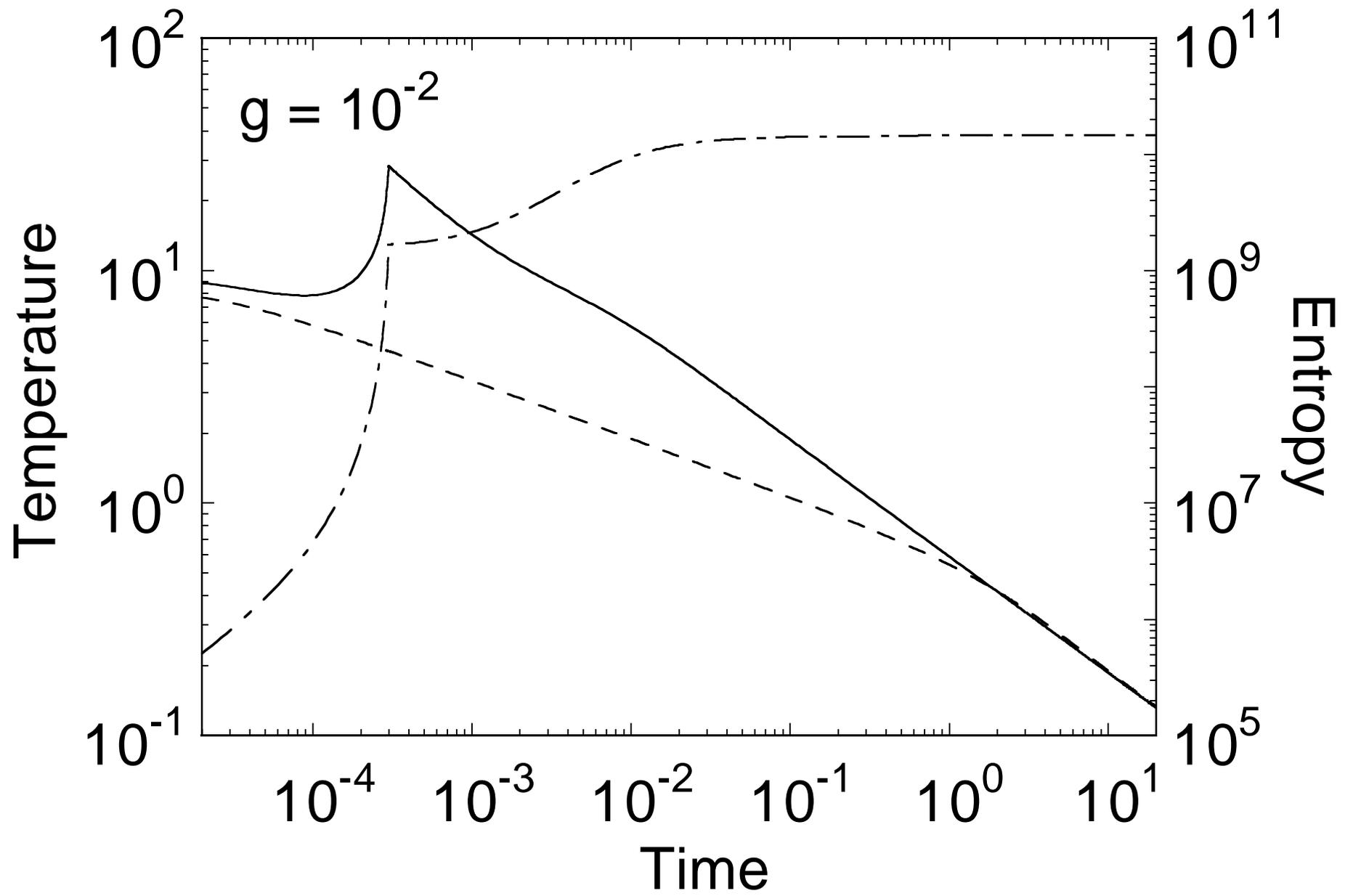
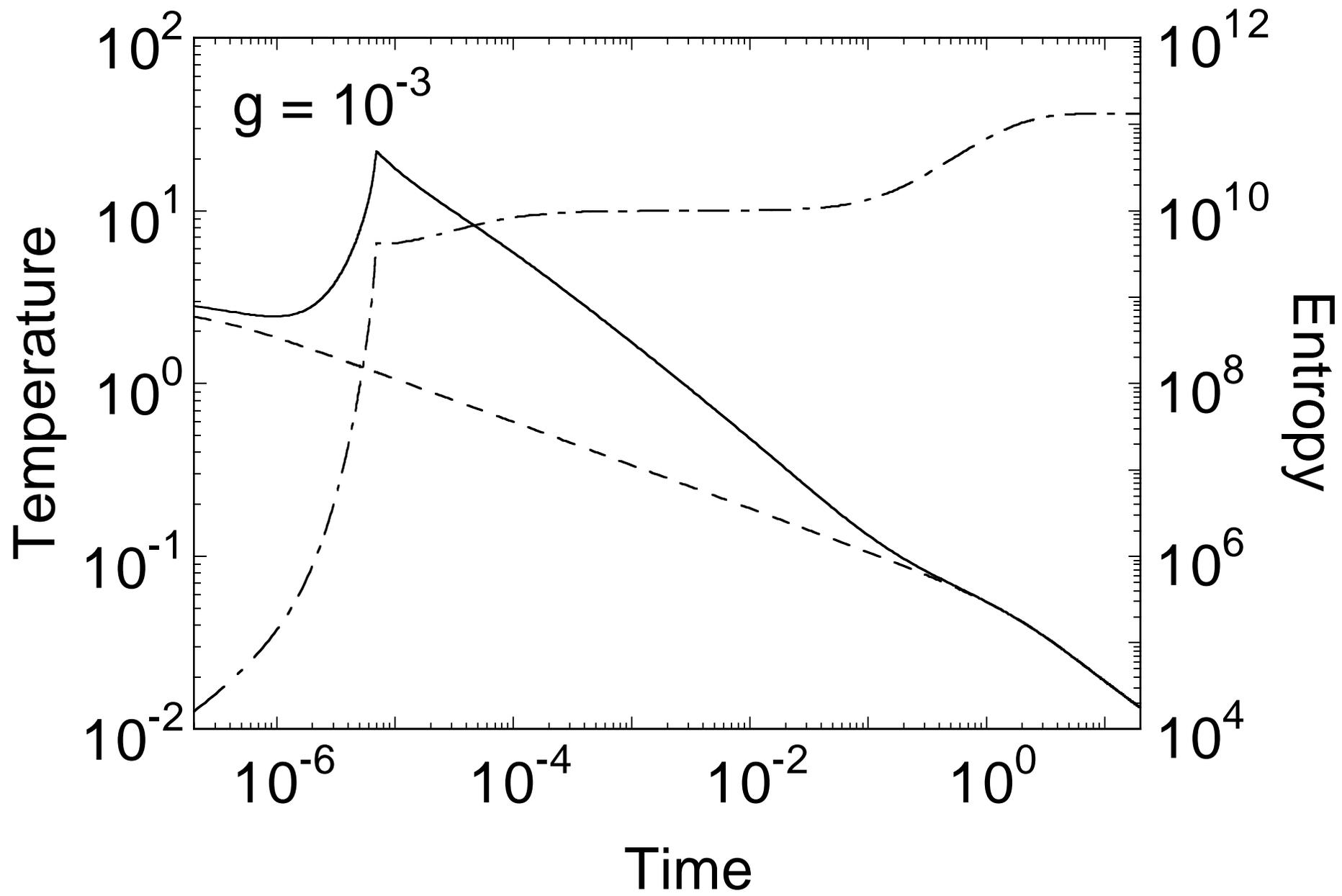
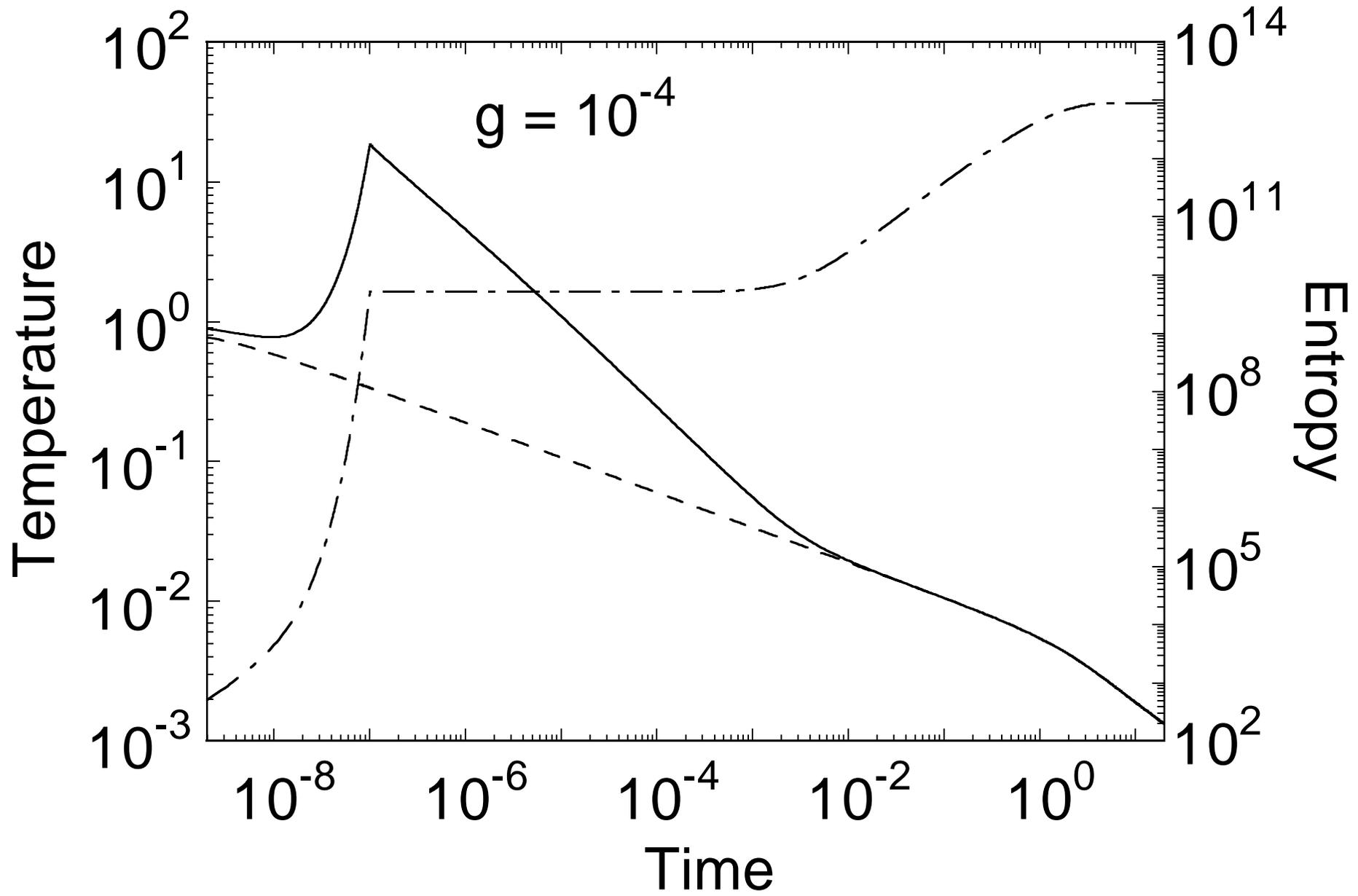
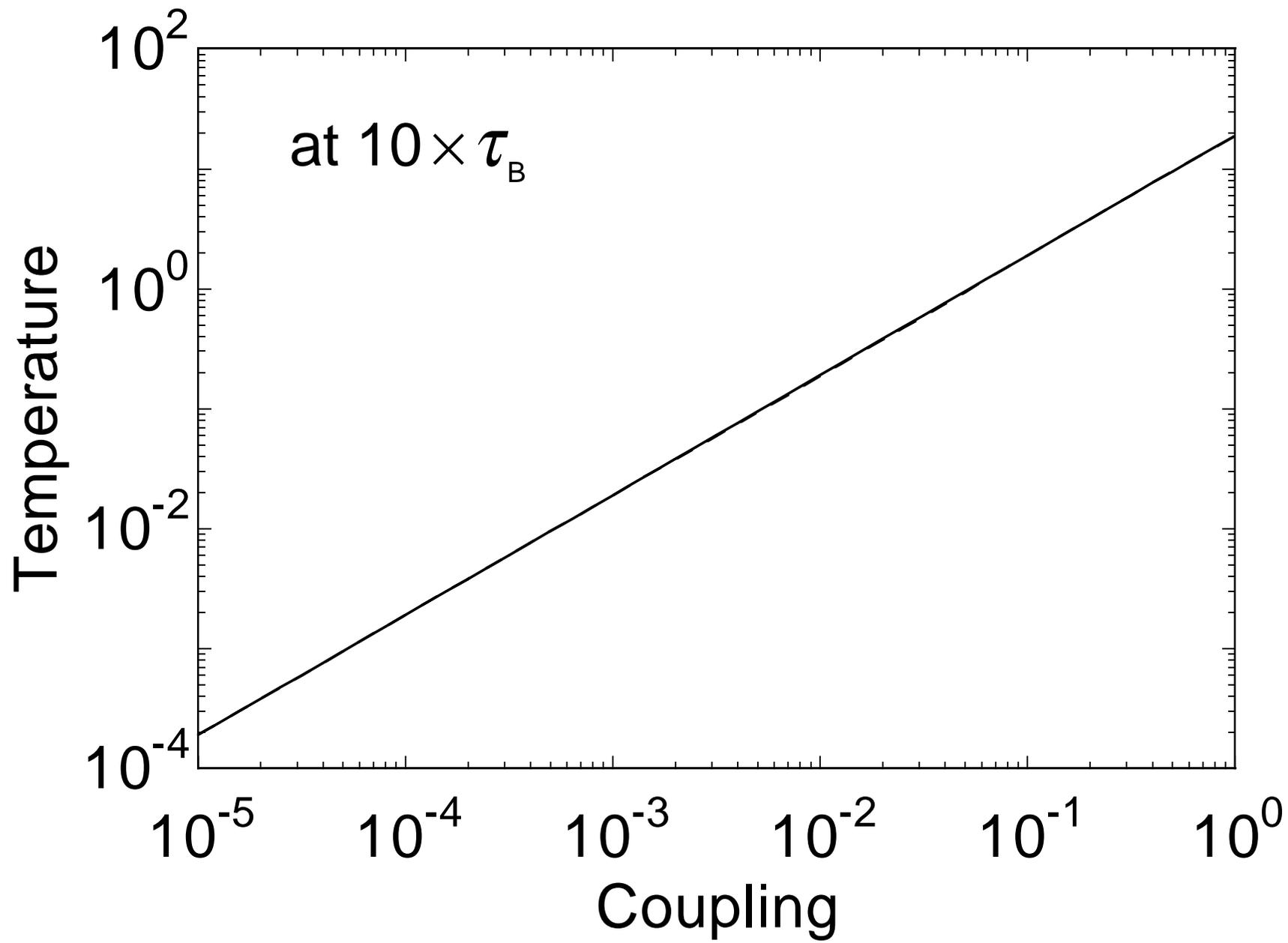
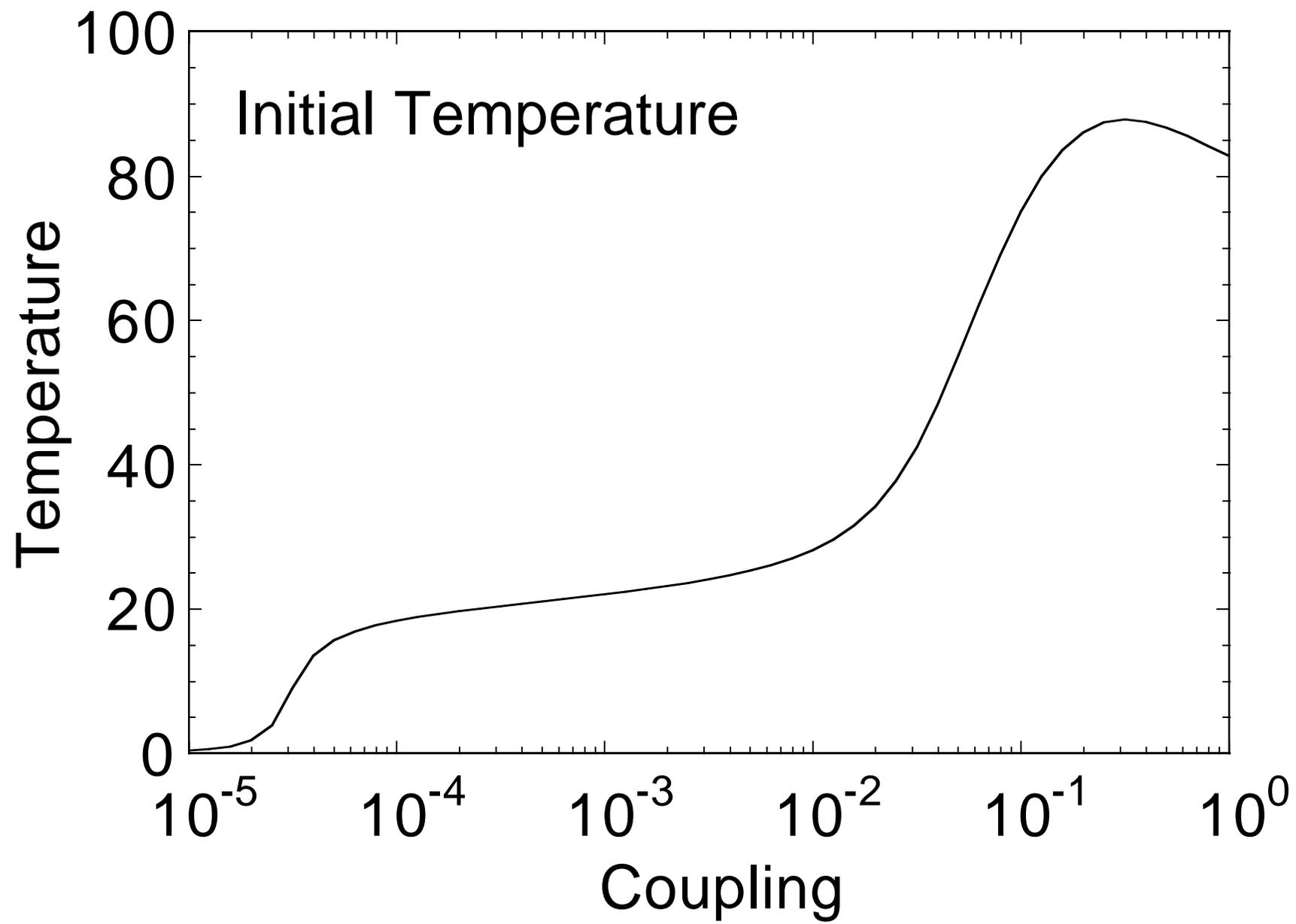
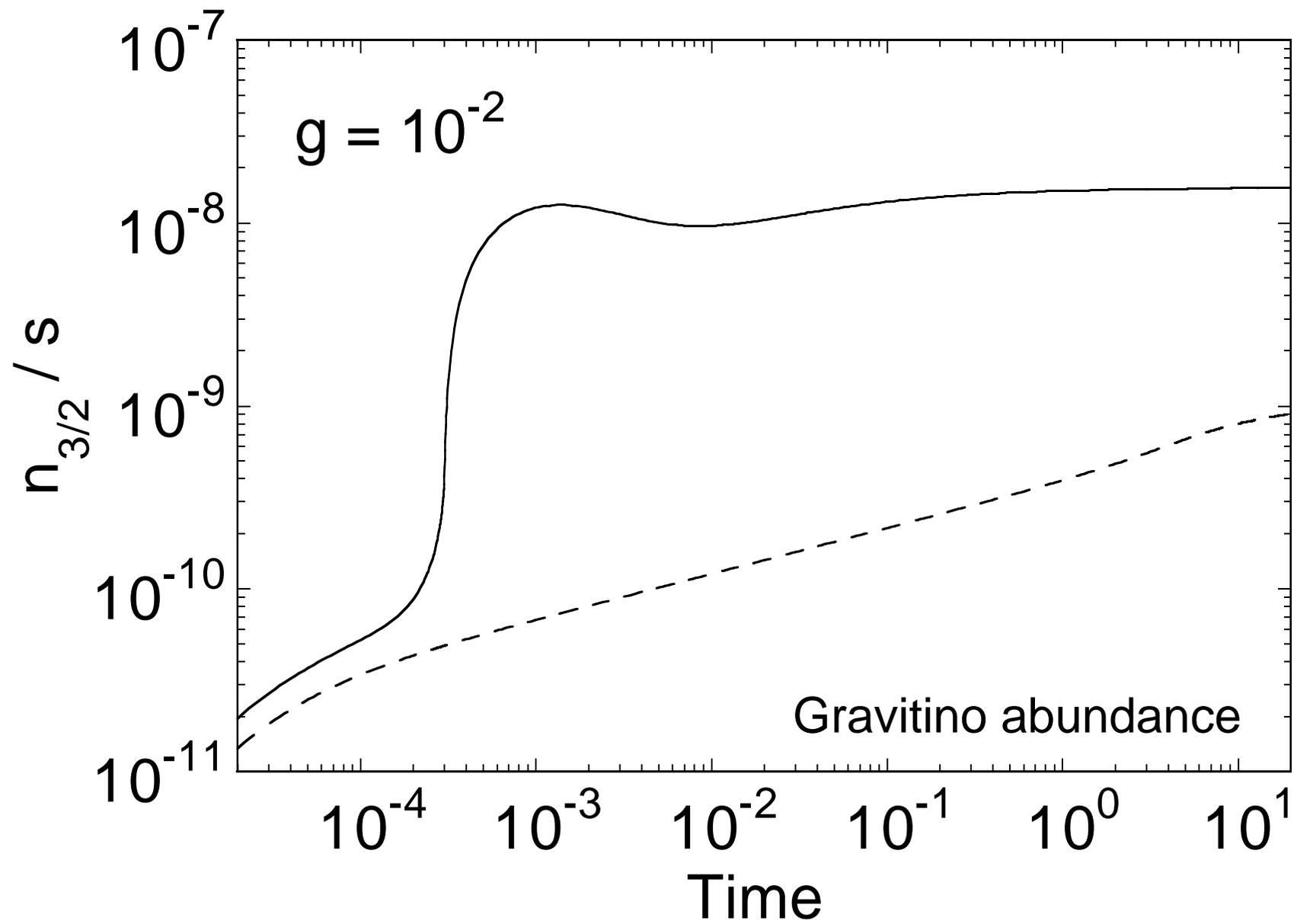
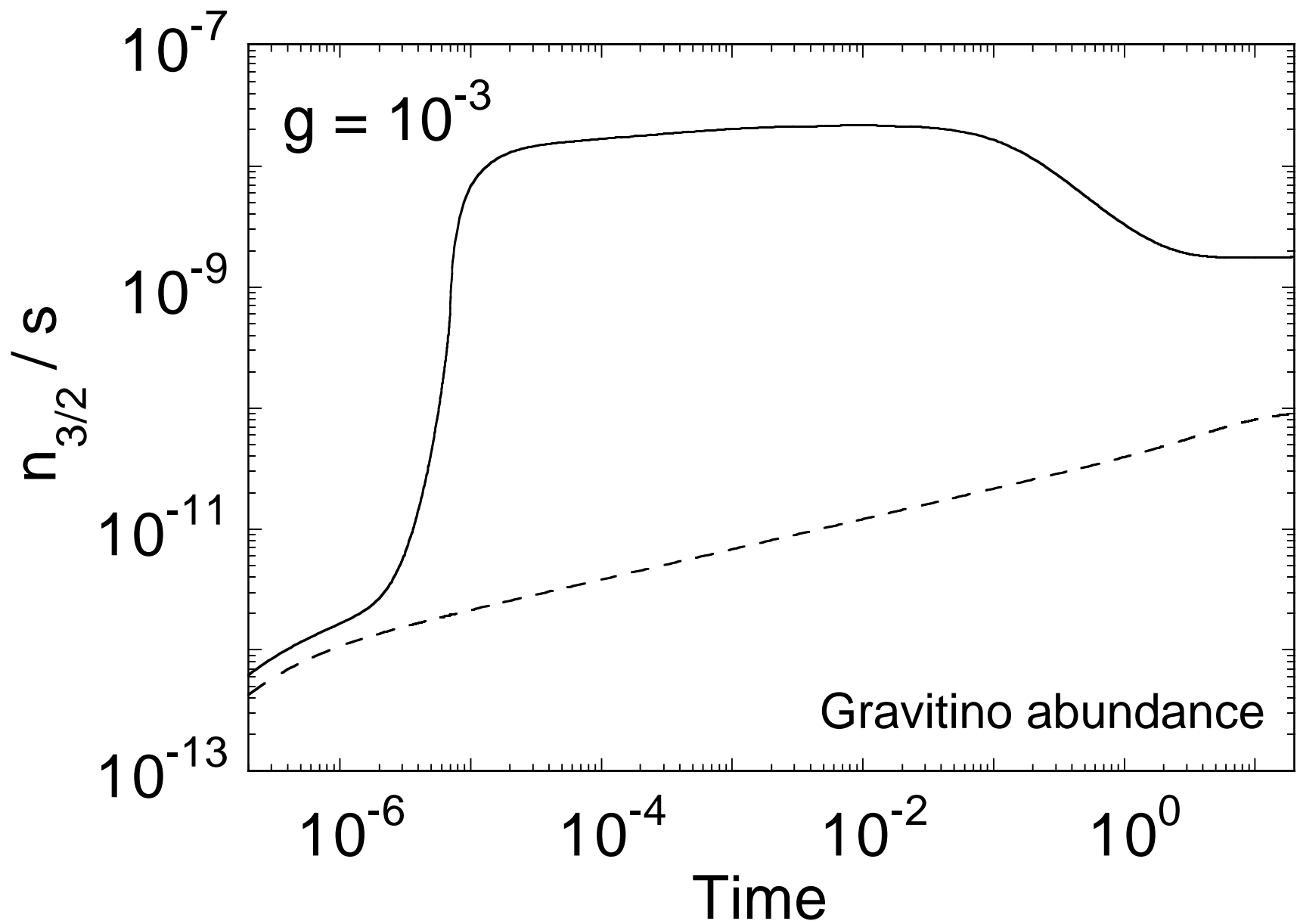
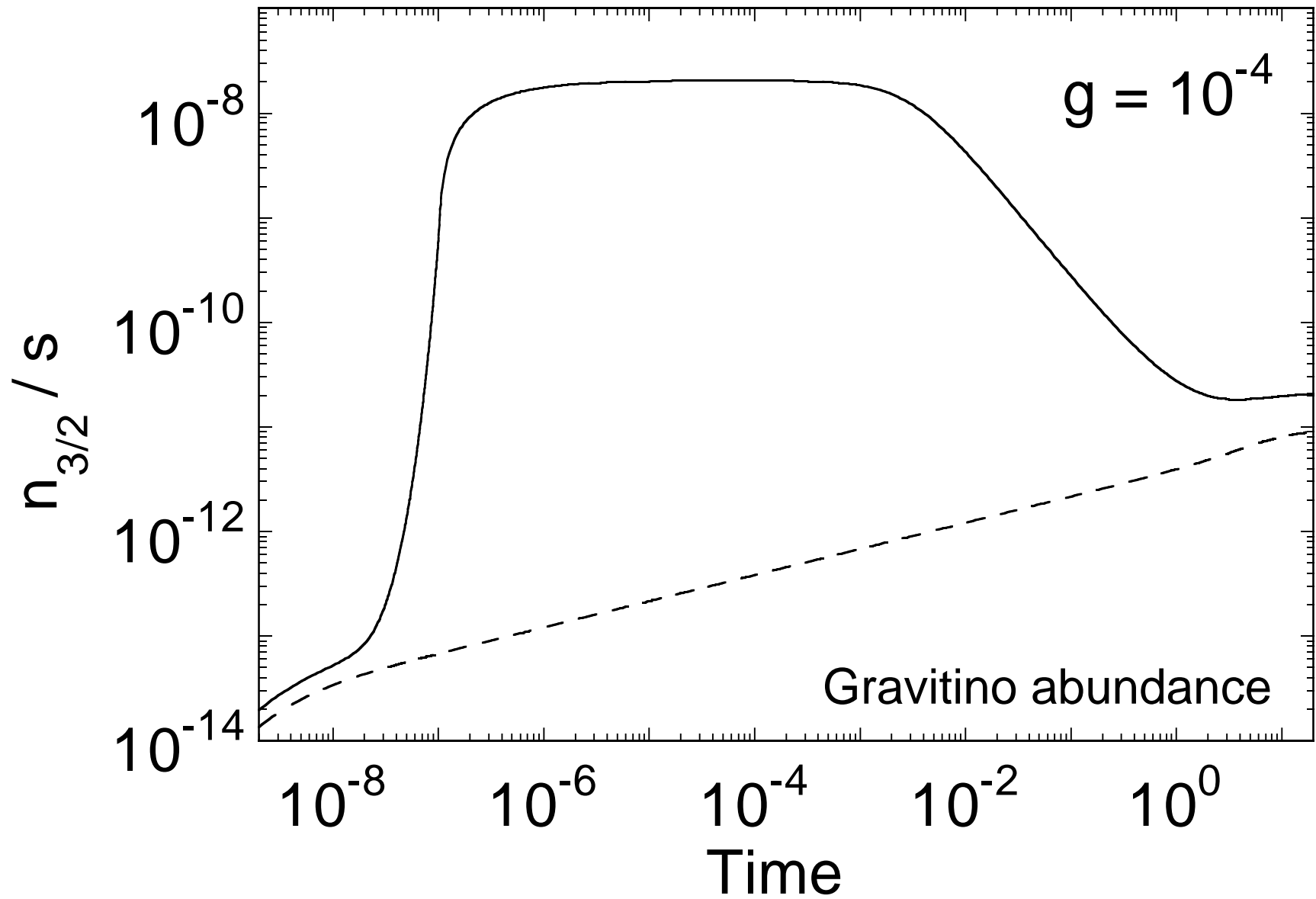
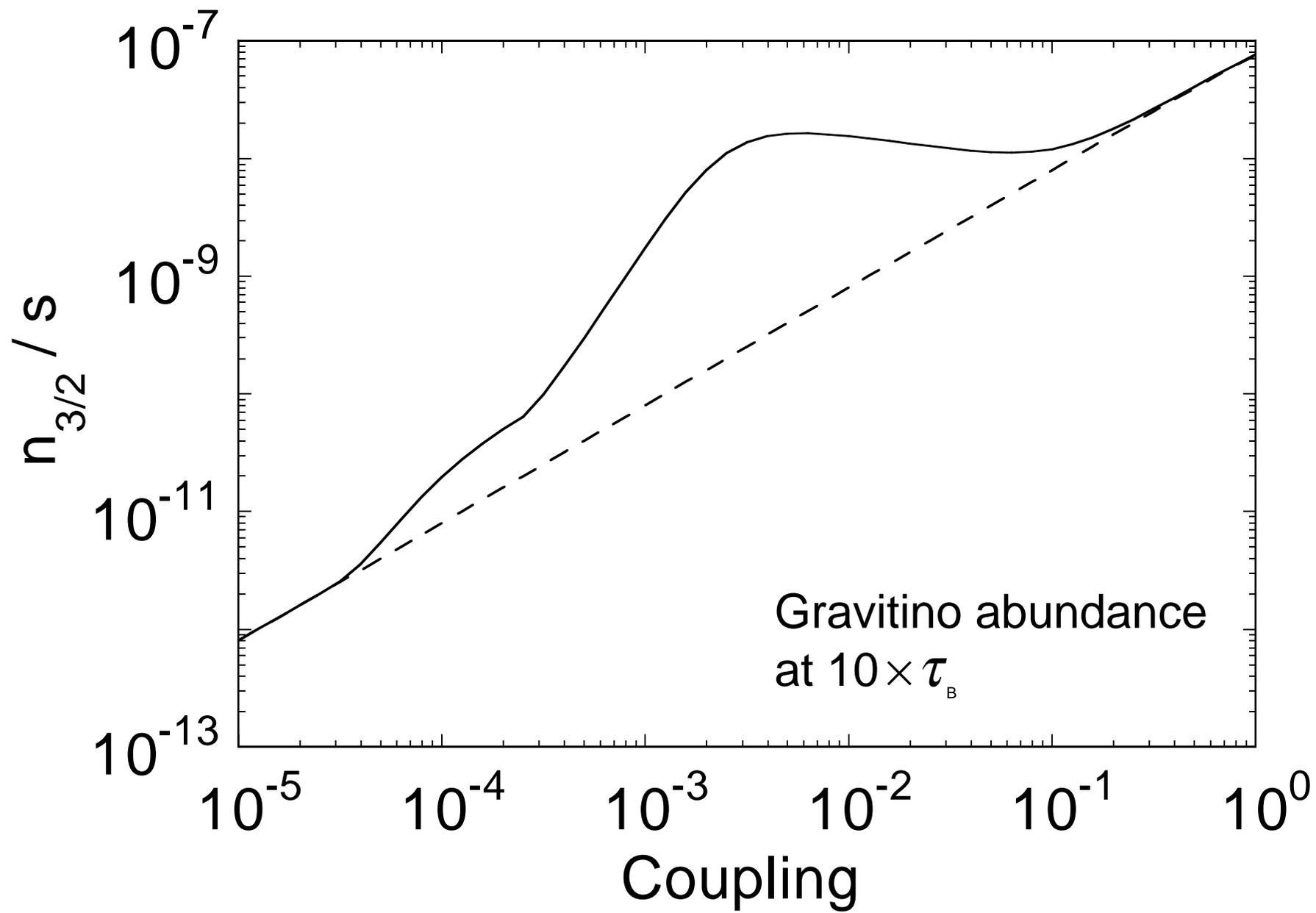
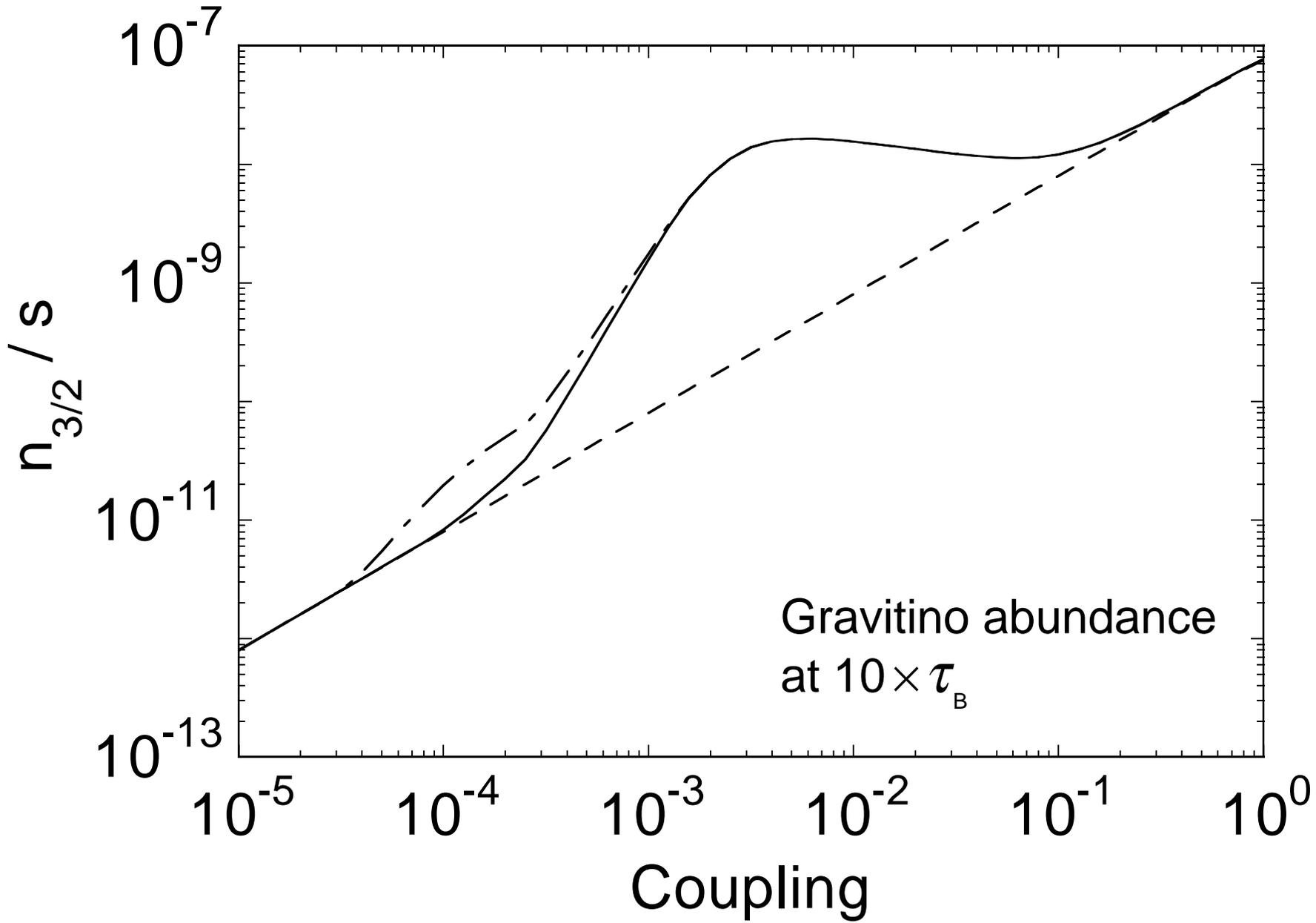
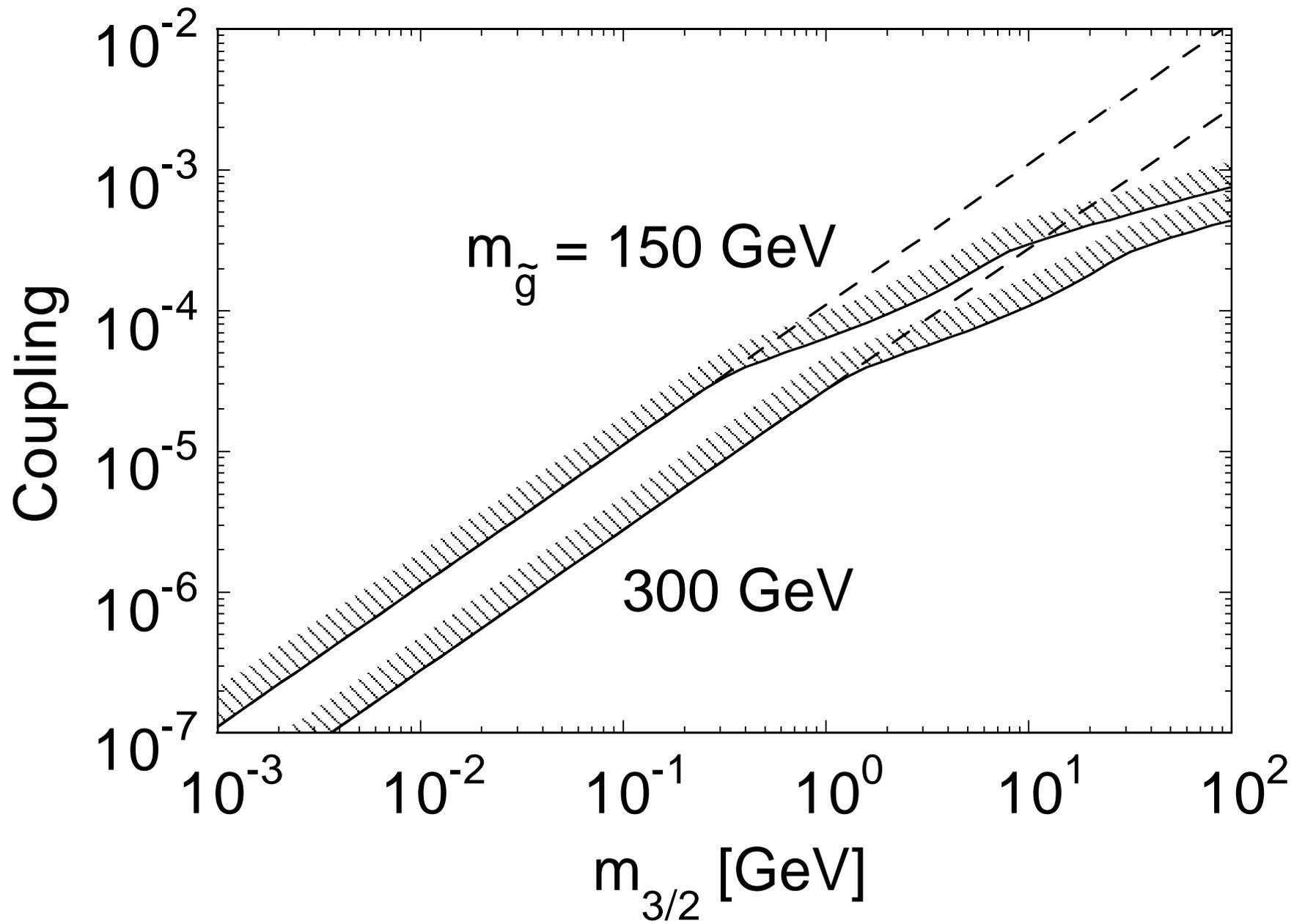
\end{document}